\documentclass[12pt]{article}
\usepackage{latexsym}
\usepackage{hyperref}
\usepackage{textcomp}

\def\NON{\nonumber\\}
\def\NXT{\\}

\def\a{\alpha}
\def\b{\beta}
\def\c{\chi}
\def\d{\delta}
\def\g{\gamma}
\def\h{\eta}

\def\j{\psi}
\def\k{\kappa}
\def\l{\lambda}
\def\m{\mu}
\def\n{\nu}

\def\p{\pi}                     % Also, \varpi
\def\th{\theta}                  %       \vartheta
\def\r{\rho}                    %       \varrho
\def\s{\sigma}                  %       \varsigma

\def\D{\Delta}

\def\L{\Lambda}
\def\O{\Omega}

\def\cf{{\cal F}}

\def\co{{\cal O}}

\def\cq{{\cal Q}}

\def\cz{{\cal Z}}
\def\cbo{{\,\raise-.15ex\Sc [\,}}                       % curly "
\def\bra#1{\Big\langle #1\Big|}                 % < |
\def\sbra#1{\left\langle #1\right|}             % variable < |
\def\sket#1{\left| #1\right\rangle}             % variable | >
\def\ddt#1{{\buildrel {\hbox{\LARGE .\kern-2pt.}} \over {#1}}}% double dot-over
\def\secteq#1{ \setcounter{equation}{0}
               \renewcommand{\theequation}{#1.\arabic{equation}} }

\def\sstyle{\scriptstyle}

\def\ie{\mbox{\it i.e.} }
\def\eg{\mbox{\it e.g.} }

\def\leqx{\,\raisebox{-1.0ex}{$\stackrel{\textstyle <}{\sim}$}\,}

\def\frac#1#2{ {\sstyle {#1\over #2} } }

\def\half{{1\over 2}}

\thicklines                         % thick straight lines for pictures (LaTeX)
\def\ssec#1#2{
  \vskip 3ex
  \noindent {\bf #1 #2}\\
  \vskip -1ex
  \hskip -4ex}

\def\mres{m_{\rm res}}
\def\mlcl{m_{\rm lcl}}
\def\mext{m_{\rm ext}}
\def\Mres{M_{\rm res}}
\def\Mext{M_{\rm ext}}
\def\Mlcl{M_{\rm lcl}}
\def\Xres{X^{\rm res}}

\def\sprod{\,{\bigcirc\kern -0.31cm s\;}\,}
\def\IbP{\;\stackrel{\circ}{=}\;}
\def\seef{{\it cf. }}

\def\tH{\tilde{H}}

\def\tl{\tilde\l}
\def\str{{\rm str}}
\def\ie{{\it i.e.}}
\def\promil{\textperthousand\ }   % from package textcomp

\long \def \blockcomment #1\endcomment{}

\begin{document}

\begin{center}
\vspace{10mm}
{\large\bf Before sailing on a domain-wall sea}
\\[12mm]
Maarten Golterman$^a$\ \ and \ \ Yigal Shamir$^b$
\\[8mm]
{\small\it
$^a$Department of Physics and Astronomy,
San Francisco State University\\
San Francisco, CA 94132, USA}\\
{\tt maarten@stars.sfsu.edu}
\\[5mm]
{\small\it $^b$School of Physics and Astronomy\\
Raymond and Beverly Sackler Faculty of Exact Sciences\\
Tel-Aviv University, Ramat~Aviv,~69978~ISRAEL}\\
{\tt shamir@post.tau.ac.il}
\\[10mm]
{ABSTRACT}
\\[2mm]
\end{center}

\begin{quotation}
We discuss the very different roles of the valence-quark and the sea-quark
residual masses ($\mres^v$ and $\mres^s$) in dynamical domain-wall fermions
simulations. Focusing on matrix elements
of the effective weak hamiltonian containing a power divergence,
we find that $\mres^v$ can be a source of a much bigger systematic error.
To keep all systematic errors due to residual masses
at the 1\% level, we estimate that one needs $a\mres^s \leqx 10^{-3}$
and  $a\mres^v \leqx 10^{-5}$, at a lattice spacing $a\sim 0.1$~fm.
The practical implications are that (1) optimal use of computer resources
calls for a mixed scheme with different domain-wall fermion actions for
the valence and sea quarks; (2) better domain-wall fermion actions are needed
for both the sea and the valence sectors.
\end{quotation}

\newpage
\vspace{5ex}
\noindent {\large\bf 1.~Introduction}
\secteq{1}
\vspace{3ex}

Lattice QCD is entering a stage in which it is becoming more
standard to include dynamical fermions (\ie\ sea quarks) in
numerical simulations, in order to move away from the quenched
approximation.  Needless to say, this step is crucial if lattice
QCD is to become really predictive, with full control of all
systematic errors.

What has not changed is that simulations with dynamical fermions
are very expensive: updating the fermion determinant takes the
bulk of the computer time.  It is therefore of key importance to
choose the parameters of any simulation such that the desired
precision of physical observables is obtained. However, in order
to use resources wisely, it is equally important to balance the
effort with respect to the various systematic errors involved.
It is questionable to spend an inordinate amount of computational
effort to control one type of error to machine precision, while
not controlling some other error to better than, for instance,
one percent.  This is especially the case if the difference
amounts to making the simulation feasible or not.

In the case of domain-wall fermions (DWF) \cite{dbk,dwf}, an important
systematic error comes from the explicit breaking of chiral symmetry
due to the fact that the size of the fifth dimension, $L_5$,
is kept finite.  Chiral symmetry is better at larger $L_5$,
but the computational cost also grows with $L_5$, making the
choice of $L_5$ an optimization problem.
Since the cost of simulating the fermion determinant is so
large, this optimization problem is particularly acute if one
wishes to simulate unquenched lattice QCD using DWF for both the valence-
and sea-quark sectors.

In this paper we propose that, given a certain
DWF discretization of the Dirac operator, the size of the fifth
dimension entering in the sea-quark Dirac operator, $L^s_5$, may be chosen
much smaller than its valence counterpart, $L^v_5$.\footnote{
  Superscripts $v$ and $s$ refer to quantities in the valence-
  and sea-quark sectors, respectively.
}
This could
be of substantial help in reducing the cost of simulations
with dynamical DWF, while keeping control over systematic
errors due to the lattice-artifact chiral symmetry breaking coming
from a finite $L_5$.

Since the quality of chiral symmetry is most important in weak
hadronic decays, we will center our discussion on the well-known
case of non-leptonic kaon decays.  In particular, we have in mind
the method, proposed long ago \cite{bernardetal}, of determining
physical kaon-decay matrix elements from simpler, unphysical ones,
such as $K^+\to\pi^+$ and $K^0\to 0$, using chiral symmetry.
As is well known, power divergences are encountered in this computation,
and good chiral symmetry is a key ingredient in controlling it.

We begin with a list of observations, on which this proposal
is based.  The rest of the paper elaborates on these observations.

%%%%%%
\vspace{3ex}\noindent
{\bf Observations}
%\vspace{-3ex}
%%%%

\begin{enumerate}

\item
Violations of chiral symmetry in Ward--Takahashi (WT) identities and
in physical matrix elements arise from the valence sector only,
essentially by definition. They vanish when approaching the valence-DWF
chiral limit $L_5^v\to\infty$, in which the ``residual quark mass"
(\ie\ the effective quark mass due to a finite $L_5$ when the
explicit quark mass has been set equal to zero)
$\mres^v$ goes to zero, regardless of the value of $\mres^s$.

\item
Consider any operator $\cq$ belonging to the effective weak hamiltonian,
transforming in $(8_L,1_R)$ of $SU(3)_L\times SU(3)_R$.
If $\mres^v=0$ then, for any $\mres^s$,
its power-divergent piece arises from a {\it single}
lower-dimension operator $\co_2$.
One subtraction, $\cq_{\rm sub}=\cq-\cf\co_2$, defined by
the renormalization condition $\sbra{0} \cq_{\rm sub} \sket{K^0} = 0$,
removes the power divergences from all
of its matrix elements simultaneously.

\item
We may constrain $\mres^v$ and $\mres^s$ by the requirement that
$\mres$-induced deviations
of physical matrix elements from the chiral limit will,
for instance, be at the 1\% level.  Since the systematic effects due to
$\mres^v\ne 0$ are magnified by inverse powers
of the lattice spacing, we end up with a rather tight constraint on
$\mres^v$.

\item
Systematic effects due to $\mres^s\ne 0$ alone are {\it not} magnified
by inverse powers of the lattice spacing when following
the subtraction procedure defined above. Therefore, chiral perturbation theory
gives rise to a milder bound on  $\mres^s$.
A comparable bound arises from considering $\mres^s$-induced
chiral-symmetry violations in the mixing among dimension-six operators
as well as scaling violations.

\end{enumerate}

Allowing for $\mres^s$ to be much less constrained than $\mres^v$
translates into the possibility of choosing $L^s_5$ much smaller
than $L^v_5$.  This, in turn, may lead to a substantial economy
of computer time, making dynamical DWF simulations possible
on currently-planned machines.

The rest of this paper is organized as follows.  In Sect.~2, we explain
the main observations (1 and 2) in more detail, paying particular attention
to power-divergent subtractions.
In Sect.~3 we analyze the situation at next-to-leading order
(NLO) in chiral perturbation theory (ChPT), with emphasis on the
role of $\mres^{v,s}$.
As this knowledge will be crucial for the bounds we intend to establish,
we give a rather detailed
account of the way power divergences relate to low-energy constants (LECs)
in ChPT to NLO,
including a discussion of short- versus long-distance effects.
Sect.~4 then gives estimates of the numerical bounds which $\mres^{v,s}$
need to satisfy in order to determine the leading-order (LO)
LECs relevant for $K\to\pi\pi$ decay to about
1\% accuracy, at an inverse lattice spacing of about 2~GeV.
In Sect.~5 we discuss the implications of our results.
The systematics of residual chiral symmetry violating
effects is discussed in Appendix~A.

%%%%%%%%%%%%
\vspace{5ex}
\noindent {\large\bf 2.~Power divergences}
\secteq{2}

%%%%%%
\ssec{2.1}{Setup and (chiral) symmetry considerations (Observation 1)}
%%%%%%
It is easiest to
illustrate the very different roles of the sea and valence chiral symmetries
in a simplified, extreme case where the valence quarks have an exact chiral
symmetry (for $m_q^v=0$), while the sea quarks are Wilson fermions,
with basically no chiral symmetry at finite lattice spacing.
This ``mixed'' framework was
studied recently in a chiral-lagrangian framework in ref.~\cite{brs}.

We will assume the number of valence quarks to be $N_v$, and the number
of sea quarks to be $N_s$.
With a mixed action, there is no (exact) symmetry that mixes the valence
and sea quarks. With Wilson sea quarks and Ginsparg--Wilson valence
(and ghost) quarks, the lattice flavor symmetries are
\begin{eqnarray}
  G_{sea} &=& U(N_s)_V
\NON
  G_{val} &=& [SU(N_v|N_v)_L \otimes SU(N_v|N_v)_R] \sprod
  U(1)_V\,.
\label{sym}
\end{eqnarray}
The $\sprod$ symbol denotes a semi-direct product.

Let us now consider an axial WT identity. It is obtained by applying
a (local) axial transformation
to the valence quarks {\it only} (and not to either sea or ghost
quarks). Since the (global) non-singlet axial symmetry
of the valence quarks is exact, there are no lattice-artifact
chiral symmetry violating terms in the resulting WT identity.
This is true in particular for the PCAC relation,
implying that the (valence-quark) pion mass must vanish exactly
for $m_v = 0$.  Whether or not the sea quarks have an axial symmetry plays
no role here!  All that is necessary is that the gauge fields
do not transform under the flavor symmetries, and therefore there is no
way to ``communicate'' (axial) flavor transformations
applied to valence quarks to the sea quarks.

%%%%%%
\ssec{2.2}{Application to $K\to\pi$ penguins (Observation 2)}
%%%%%%
Similar considerations apply to any operator mixing which occurs in
the calculation of physical matrix elements. The basic reason is that,
by definition, the operators that
interpolate the external states are made of valence quarks only.
This can be best illustrated through an example.

In the calculation of matrix elements needed for $K\to \p\p$,
the most dangerous mixings are the power-divergent subtractions.
This allows the four-fermion operators occurring in those matrix elements
to mix with the fermion bilinear  \cite{bernardetal}
\begin{eqnarray}
  \co_2
  &=&
  \bar{s}\g_\mu(\overrightarrow{D}_\mu-\overleftarrow{D}_\mu)(1-\g_5)d
\NON
  &\IbP&
  (m_s+m_d)\bar{s}d + (m_d-m_s)\bar{s}\g_5 d \,,
\label{O4}
\end{eqnarray}
where $\IbP$ denotes on-shell equality.\footnote{In this paper, we only
consider matrix elements with on-shell external states, and we may thus
use the equations of motion.  However, we will allow for matrix elements
which do not conserve momentum.}  In full continuum QCD
the operators that can mix with a fermion bilinear are the $(8_L,1_R)$ ones.
In a hybrid scheme, the four-fermion operator must contain valence $\bar{s}$
and $d$. The other two fermions may be valence, sea, or ghost quarks.
When the valence sector has the exact chiral symmetry of
Eq.~(\ref{sym}) with $N_v=3$, the effective weak hamiltonian
contains four-fermion operators which transform as $(8_L,1_R)$ under
$SU(3)_L\times SU(3)_R\subset G_{val}$.
They can thus mix with an $\bar{s}d$ bilinear with the same
quantum numbers (including $CPS$ symmetry \cite{bernardetal}, under
which the weak hamiltonian
is even).\footnote{In the case of the strong penguins $Q_{5,6}$ only
the ``partially-quenched singlet (PQS)" operators of ref.~\cite{gppeng}
can mix with $\bar{s}d$. In this paper, we consider only PQS operators.}
In order to be able to use the $G_{val}$
symmetry in constructing such bilinears, the quark mass matrix $M$ is
promoted to a spurion field, transforming as $M\to LMR^\dagger$,
with $L,R\in SU(N_v|N_v)_{L,R}$.  The only
fermion bilinear with dimension smaller than six
satisfying these requirements is $\co_2$.
This operator has mass-dimension four, and hence it mixes with
the octet weak hamiltonian through a coefficient proportional to
$1/a^2$.  It also follows that there is no mixing of order $1/a$,
because a dimension-five operator with the appropriate quantum
numbers does not exist.

%%%%%%
\ssec{2.3}{Renormalization condition for the power divergence}
%%%%%%
In field theory one renormalizes operators,
not individual matrix elements.\footnote{For a similar
discussion in this context, see Ref.~\cite{RBC}.}
Given a four-fermion operator $\cq$ which mixes with $\co_2$,
we thus need to define a subtracted operator
\begin{equation}
  \cq_{\rm sub} = \cq - \cf\, \co_2 \,.
\label{subt}
\end{equation}
Here $\cf$ is a dimensionful function of the parameters of the theory
which diverges like $1/a^2$.
Among these parameters are the explicit quark masses contained in the
lattice action. If we assign to the mass parameters their spurion
transformation properties, $\cf$ must be invariant under all symmetries
(for the hybrid theory,
this includes in particular the symmetries of Eq.~(\ref{sym}) and $CPS$).

As already mentioned, no lower-dimension operator other than
$\co_2$ has the quantum numbers of $\cq$. (This is not true if
$\mres^v \ne 0$, see below.) Hence, a single renormalization
condition is sufficient to determine the subtraction.
Moreover, the single subtraction~(\ref{subt}) must remove the entire
power-divergent part of $\cq$ (see below for logarithmic divergences).
Here we choose the homogeneous renormalization condition\footnote{
This renormalization condition was also discussed in the context
of the direct determination of $K\to\pi\pi$ matrix elements in
ref.~\cite{dawsonetal}.}
\begin{equation}
  0 = \bra{0} \cq_{\rm sub} \sket{K^0}
  =  \bra{0} \cq \sket{K^0} - \cf\,(m_d-m_s)
     \bra{0} \bar{s}\g_5 d \sket{K^0}\,,
\label{Kvac}
\end{equation}
which yields
\begin{equation}
  \cf = {\sbra{0} \cq \sket{K^0}
        \over (m_d-m_s)\sbra{0} \bar{s}\g_5 d \sket{K^0}  } \,.
\label{F}
\end{equation}
Equation~(\ref{Kvac}) is a consistent renormalization condition because,
for $m_d \ne m_s$, the power-divergent part of $\cq$ contributes
to the matrix element at LO in ChPT.
Indeed, the power divergence corresponds to
(the matrix element of) the operator $\bar{s}\g_5 d$.
This operator is a total divergence on shell, which contributes to $K^0 \to 0$
because momentum is not conserved in that matrix element
\cite{bernardetal}.  Since momentum is conserved for the physical
$K\to\pi\pi$ matrix elements, one does have that $\sbra{\pi\pi}\co_2
\sket{K}=0$.

Before we move on, we need to discuss the freedom in choosing the
renormalization condition. Generically, a physical renormalization condition
may be written as $\sbra{A} \cq_{\rm sub} \sket{B} = C$, where
$\sket{A}$ and $\sket{B}$ are some states, and the (not necessarily zero)
constant $C$ is finite in physical units. How does changing the renormalization
condition affect the subtracted operator and its matrix elements?
Let us expand the subtraction coefficient in Eq.~(\ref{subt})
as a power series in $a$:
\begin{equation}
  \cf(a) = {\cf_2\over a^2}
  + \cf_0 +O(a) \,.
\label{Fa}
\end{equation}
If we choose a different renormalization condition, we will end
up with a new subtraction coefficient
\begin{equation}
  \cf'(a) = {\cf'_2\over a^2}
  + \cf'_0 + O(a) \,.
\label{Fa'}
\end{equation}
How are $\cf(a)$ and $\cf'(a)$ related?
Let $\cq'_{\rm sub} = \cq - \cf' \co_2$. Then
\begin{eqnarray}
  \bra{0} \cq'_{\rm sub} \sket{K^0}
  &=& \bra{0} \cq \sket{K^0}
      - \left({\cf'_2\over a^2}
      + \cf'_0 + O(a) \right)
      (m_d-m_s)\bra{0} \bar{s}\g_5 d  \sket{K^0}
\label{FF'}
\NXT
  &=& \left({\cf_2-\cf'_2\over a^2}
  + \cf_0-\cf'_0
      + O(a) \right)
      (m_d-m_s)\bra{0} \bar{s}\g_5 d  \sket{K^0}\,.
\nonumber
\end{eqnarray}
The last equality is true because
$\sbra{0} \cq_{\rm sub} \sket{K^0}=0$ by construction.
We see that $ \sbra{0} \cq'_{\rm sub} \sket{K^0}$ is finite
{\it if and only if} $\cf_2=\cf'_2$.
Now, since $\cq$ can mix with only one lower-dimension operator,
it follows that any consistent renormalization condition must
give rise to the same value for $\cf_2$.
In other words, the power-divergent part of $\cf$ is {\it independent}
of the renormalization condition, but the finite part is not.
(That is, in general
$\cf_0 \ne \cf'_0$. Of course, the same is true for the $O(a)$ part.)
In the case at hand, the subtraction defined by Eq.~(\ref{subt})
and the renormalization condition $\sbra{0} \cq_{\rm sub} \sket{K^0}=0$
ensures the absence of power divergences
in $\sbra{\p} \cq_{\rm sub} \sket{K}$
and $\sbra{\p\p} \cq_{\rm sub} \sket{K}$ simultaneously.
Denoting the coefficient of $1/a^2$ in the divergent part of a matrix element
by ``$div$,'' this statement can be summarized in
the suggestive form
\begin{equation}
  {div\sbra{A}\cq\sket{B}\over \sbra{A}\co_2\sket{B}}
  =
  {div\sbra{0}\cq\sket{K^0}\over \sbra{0}\co_2\sket{K^0}}
  =\cf_2\,,
\label{ratio}
\end{equation}
for any two states $\sbra{A}$ and $\sket{B}$.

The matrix elements of $\cq_{\rm sub}$ will in general still contain
logarithmic divergences. Fully renormalized operators may be defined
on the lattice using \eg the regularization independent (RI)
non-perturbative renormalization (NPR) method
\cite{martinellietal,RBC}. (For a different method, see ref.~\cite{CPPACS}.)
Renormalized operators
$\cq_r^i = \cz^i_j \cq^j_{\rm sub}$ are defined by the requirement
that their matrix element between given (gauge fixed, off shell)
high-momentum four-quark states has a prescribed value.
Typically, one performs a second (finite) renormalization in order to relate
the lattice-renormalized operators to renormalized operators defined
in some continuum scheme such as $\overline{\rm MS}$.

The entire renormalization procedure discussed above does not rely
on the effective chiral lagrangian.
The numerical extraction of $\cf^j$ can be done directly from Eq.~(\ref{Kvac}).
Similarly, the NPR matrix $\cz^i_j$ is determined numerically.
In the case at hand, we
obtain renormalized $K\to\pi$ matrix elements
(ignoring for simplicity the renormalization of the external states) via
\begin{equation}
  \sbra{\p} \cq_r^i \sket{K}
  = \cz^i_j \sbra{\p} \cq_{\rm sub}^j \sket{K}
  = \cz^i_j \left( \sbra{\p} \cq^j \sket{K} - \cf^j (m_d+m_s)
     \sbra{\p} \bar{s}d \sket{K} \right)\,.
\label{Kpi}
\end{equation}
This result can now be matched on to the appropriate
ChPT expression. The LECs obtained this way are then used
to determine the $K\to\p\p$ matrix elements.

While its is clear that the complete renormalization is carried out
without using ChPT, the ChPT calculation must conform to the actual
renormalization procedure that we envisage performing numerically.
We will return to this point below.

%%%%%%%%%%%%
\vspace{5ex}
\noindent {\large\bf 3.~Chiral perturbation theory}
\secteq{3}
\vspace{3ex}

In this section we work out a ChPT example.
The central point is that, as long as $\mres^v=0$, the above
procedure completely removes the power-divergences
even if both the un-subtracted $\cq$ and $\co_2$ matrix elements
depend on $\mres^s$.

%%%%%%
\ssec{3.1}{Matrix elements of $\cq$ and $\co_2$ for $\mres=0$}
%%%%%%
We begin with the representation of a general weak $(8_L,1_R)$ operator
to NLO in ChPT in the continuum, working
in the limit in which $\mres^{v,s}=0$.
In this case, the theory is partially quenched \cite{bg}, with an
enlarged symmetry group\footnote{For our purposes, it is sufficient
to consider this symmetry group, even though the actual symmetry group
is slightly different, see Refs.~\cite{truesym}.}
\begin{equation}
[SU(N_v+N_s|N_v)_L\times SU(N_v+N_s|N_v)_R]\times U(1)_V\,.\label{pqsym}
\end{equation}
Any octet operator $\cq$ mediating $\Delta S=1$ transitions can be written as
(in euclidean
space)\footnote{We will use the notations and conventions of
ref.~\cite{gp}.}
\begin{eqnarray}
\cq&=&\cq_2+\cq_4+\dots\,,
\label{lweak}\\
\cq_2&=&-\a^8_1\,\str(\L L_\mu L_\mu)+\a^8_2\,\str(\L X_+)\,,
\NON
\cq_4&=&{1\over (4\pi f)^2}\left(\sum_i\b^8_i\co^8_i
+\sum_i\tilde{\b}^8_i\tilde{\co}^8_i\right)\,,\nonumber
\end{eqnarray}
in which $\a^8_i$ and $\b^8_i$ denote LECs
at LO and NLO, respectively, and
\begin{equation}
\L^i_j=\d_{i3}\d_{j2}\,.\label{Lambda}
\end{equation}
In terms of the non-linear
field $\Sigma=\exp{(2i\Phi/f)}$,
\begin{eqnarray}
L_\mu&=&i\Sigma\partial_\mu\Sigma^\dagger\,,\label{LandX}\\
X_\pm&=&2B_0(\Sigma M^\dagger\pm M\Sigma^\dagger)\,.\nonumber
\end{eqnarray}
$M$ is the (diagonal) quark mass matrix containing the valence,
sea and ghost quark masses and, after
promotion to a spurion field, transforming just like
$\Sigma$ under the full chiral group $G_L\times G_R$,
$\Sigma\to L\Sigma R^\dagger$, with $L,R\in G_{L,R}$.
The list of NLO operators relevant for our purposes is \cite{kmw}
\begin{eqnarray}
\co^8_1&=&\str(\L X_+X_+)\,,\label{nloops}\\
\co^8_2&=&\str(\L X_+)\,\str(X_+)\,,\NON
\co^8_3&=&\str(\L X_-X_-)\,,\NON
\co^8_4&=&\str(\L X_-)\,\str(X_-)\,,\NON
\co^8_5&=&\str(\L [X_+,X_-])\,,\NON
\co^8_{10}&=&\str(\L \{X_+,L_\mu L_\mu\})\,,\NON
\co^8_{11}&=&\str(\L L_\mu X_+L_\mu)\,,\NON
\co^8_{13}&=&\str(\L X_+)\,\str(L_\mu L_\mu)\,,\NON
\co^8_{14}&=&\str(\L L_\mu L_\mu)\,\str(X_+)\,,\NON
\co^8_{15}&=&\str(\L [X_-,L_\mu L_\mu])\,. \nonumber
\end{eqnarray}
There are two total-derivative operators relevant for our discussion
\cite{gp},
\begin{equation}
\tilde{\co}^8_1=i\partial_\mu\,\str(\L[L_\mu,X_+])\,,\ \ \
\tilde{\co}^8_2=i\partial_\mu\,\str(\L\{L_\mu,X_-\})\,.
\label{tdop}
\end{equation}
In partially quenched QCD, $O^8_{14}$ is an independent operator,
which cannot be written in terms of the other operators \cite{ls}.%
\footnote{A point which has been overlooked in ref.~\cite{gp}.}
However, the operators $\co^8_1$, $\co^8_3$ and $\co^8_5$ are
not independent:
\begin{equation}
\co^8_1-\co^8_3+\co^8_5=4\,\str(\L MM^\dagger)=0\,,\label{depop}
\end{equation}
where the latter equality holds because $M$ is diagonal.
Henceforth, we will set $\b^8_5=0$.

We also need the ChPT representation of the operator $\co_2$
to NLO.  This operator can be written as the $\Delta S=1$ part
of the variation of the (continuum) QCD lagrangian under a
chiral rotation which transforms $M\to(\L+\L^\dagger)M$.  Since the QCD
lagrangian is represented in ChPT by the strong chiral lagrangian,
all we need to do is to perform this same chiral rotation on
the strong chiral lagrangian.  Doing this, keeping only the
$\L$ ($\Delta S=+1$) part, one obtains \cite{kmw}
\begin{eqnarray}
\co_2&=&{f^2\over 8}\,\str(\L X_+)
\label{o2chpt}\\&&+
(L_8+{1\over 2}H_2)\co^8_1+2L_6\co^8_2+(L_8-
{1\over 2}H_2)\co^8_3+2L_7\co^8_4+{1\over 2}H_2\co^8_5
\NON
&&+{1\over 2}L_5\co^8_{10}+L_4\co^8_{13}-{1\over 2}L_5\co^8_{15}
+\tilde{L}_1(\tilde{\co}^8_1+\tilde{\co}^8_2)\,,
\nonumber
\end{eqnarray}
where $L_i$ and $H_2$ are $O(p^4)$ Gasser--Leutwyler coefficients
\cite{gl}; $\tilde{\co}^8_1+\tilde{\co}^8_2$ is the chiral rotation of
$2i\partial_\mu\,\str(L_\mu X_-)$,
and $\tilde{L}_1$ is a new strong LEC multiplying this operator.
We have added this total-derivative
operator just as we did in Eq.~(\ref{lweak}).  (The operator $\co_2$ is a local
operator, not integrated over space-time.  So, more precisely,
one may obtain the operator $\co_2$ by doing the chiral rotation
on the strong energy-momentum tensor, which
may contain total-derivative terms.)
We may set $H_2=0$ because of Eq.~(\ref{depop}).

It is now straightforward to calculate $\cf$, defined in Eq.~(\ref{F}),
to NLO in ChPT, using the results for $\sbra{0}\cq\sket{K^0}$ from
ref.~\cite{gp}, and Eq.~(\ref{o2chpt}) above.  Since at LO only $\str(\L X_+)$
contributes to $K^0\to 0$, the chiral
logarithms proportional to $\a^8_2$ drop out of $\cf$, and we find
\begin{eqnarray}
\cf&=&{8\over f^2}\Biggl[\a^8_2+\a^8_1\times({\rm chiral\ logs})
\label{Fchpt}\\
&&\hspace{-0.6cm} +\; {B_0\over(4\pi f)^2}
\Biggl((4\b^8_1-32\l_8\a^8_2+2\tilde{\b}^8_1-16\tilde{\l}^8_1\a^8_2)(m_s+m_d)
+(4\b^8_2-64\l_6\a^8_2)\sum_i m^s_i\Biggr)\Biggr]\,,\nonumber
\end{eqnarray}
in which
\begin{equation}
\l_i=16\pi^2L_i\,.\label{lL}
\end{equation}

We may now use this result for $\cf$ to define the subtracted operator
$\cq_{\rm sub}$ in ChPT, and calculate its $K^+\to\pi^+$ matrix element
for degenerate valence quark masses $m^v$.
We find that the LO terms proportional to $\a^8_2$,
as well as the chiral logarithms generated at NLO by the corresponding
operators, cancel in the matrix elements
$\sbra{\pi^+}\cq\sket{K^+}$ and $\cf\sbra{\pi^+}\co_2\sket{K^+}$,
leaving us with the result
\begin{eqnarray}
\sbra{\pi^+}\cq_{\rm sub}\sket{K^+}&=&{8B_0m^v\over f^2}\Biggl[
\a^8_1(1+{\rm chiral~logs})\label{ktopi}\\
&&\hspace{-3.1cm}
-\; {4B_0\over(4\pi f)^2}\Biggl(\left(2\b^8_3+2\b^8_{10}+\b^8_{11}
+8\l_5(\a^8_1-\a^8_2)-16\l_8(\a^8_1+\a^8_2)-\tilde{\b}^8_1+
8\tilde{\l}^8_1\a^8_2\right)m^v\NON
&&\hspace{-3.1cm}
+\left(\b^8_{14}-16\l_6\a^8_1+8\l_4\a^8_1\right)\sum_i m^s_i\Biggr)
\Biggr]\,,\nonumber
\end{eqnarray}
where we took the limit $m_s\to m_d=m^v$ in expression (\ref{Fchpt})
for $\cf$.

The ``chiral logs"
are those coming from the $\a^8_1$ operator in Eq.~(\ref{lweak}), and
an explicit expression can be found in ref.~\cite{gp}.\footnote{See
Eq.~(4.19) of that paper.}
The NLO terms
proportional to $\a^8_1$ come from the fact that we expressed the
common mass $M_{phys}$ of the physical (valence) meson,
which appears at LO in the $\a^8_1$ term, in
terms of the valence quark mass $m^v$ and the sea quark masses $m^s_i$, using
\begin{equation}
\left(M_{phys}\right)^2\!=\!2B_0 m^v\left(1+{\rm chiral~logs}+
{32B_0\over (4\pi f)^2}
\left[(2\l_8-\l_5)m^v+(2\l_6-\l_4)\sum_i m^s_i\right]\right)\,.
\label{massren}
\end{equation}
Finally, we omitted wave-function
renormalizations for the states $\sket{K^+}$ and $\sket{\pi^+}$.

Let us discuss what we learn from these results.
In the NLO part of Eq.~(\ref{ktopi}), $m^v$ and $\sum_i m^s_i$ are
multiplied by linear combinations of LECs having the generic form
\begin{equation}
  \sum_i c_i \b^8_i - \a^8_2 \sum_j d_j \l_j \,,
\label{comb}
\end{equation}
where $c_i$ and $d_j$ are numerical coefficients.
We know from the general arguments of Sect.~2 that
the subtracted matrix element in Eq.~(\ref{ktopi}) contains no power divergence.
Since $m^v$ and $m^s_i$ are free parameters, it follows that the
linear combinations~(\ref{comb}) are not power divergent either!
This simple observation leads to several important conclusions.

\vspace{1ex}\noindent
{\it Corollary 1.} Some of the LECs $\b^8_i$ diverge like $1/a^2$, just like
$\a^8_2$.
{\it Proof:}
The linear combination
$\sum_j d_j \l_j$ occurring in expression~(\ref{comb}) will in general
be non-zero. Since $\a^8_2$ diverges like $1/a^2$, so must $\sum_i c_i \b_i$.
(This observation is implicit in the work of ref.~\cite{ls}.)

\vspace{1ex}\noindent
{\it Corollary 2.} At NLO, un-subtracted matrix elements contain contributions
that diverge like $(1/a^2)(B_0 m/f^2)^n$ for both $n=1$ and $n=2$.
(Here $m$ stands for a generic mass parameter.)
{\it Proof:} For $n=1$ this is evident from the LO result.
For $n=2$, observe that whenever $\co^8_i$ occurs in the matrix element
$\sbra{\pi^+}\co_2\sket{K^+}$ with a non-zero coefficient,
the corresponding (power divergent!) LEC $\b^8_i$  occurs
in $\sbra{\pi^+}\cq\sket{K^+}$, where it is
multiplied by $(B_0 m/f^2)^2$.
It is quite evident that, as we go to higher
orders in ChPT, un-subtracted matrix elements will contain
divergent pieces $(1/a^2)(B_0 m/f^2)^n$ with correspondingly higher
values of $n$.

\vspace{1ex}
The last observation may seem surprising at first, because intuitively
one expects that short-distance divergences should be insensitive to
long-distance effects. We believe that this intuitive notion is, in fact,
correct, but it has to be stated more carefully.
This leads us to the following two conjectures.

\vspace{1ex}\noindent
{\it Conjecture 1.} For any operator $\co^8_i$ occurring in the NLO
representation of $\co_2$, the linear combination
$\b_i^{\rm sub} = \b_i - 8\a^8_2 \sum_j e_{ij}\l_j$
contains no power divergence. Here the linear combination $\sum_j e_{ij}\l_j$
is equal to the combination of strong LECs that multiplies $\co^8_i$
in Eq.~(\ref{o2chpt}),\footnote{
By inspection, for each $i$, $e_{ij}$ is non-zero only for one value of $j$.}
 with $L_j \to \l_j$.\footnote{This claim is already contained
in ref.~\cite{ls}, see in particular Table~5 therein.}

It is easy to understand why the combinations
$\b_i^{\rm sub}$ should be finite.
Seven of the NLO operators $\co^8_i$ in Eq.~(\ref{nloops})
as well as the two total-derivative operators in Eq.~(\ref{tdop})
occur in the representation of $\co_2$ (Eq.~(\ref{o2chpt})).
Separately, the matrix elements $\sbra{A}\cq\sket{B}$
and  $\cf\sbra{A}\co_2\sket{B}$ each
diverge like $1/a^2$. But taken together they yield
$\sbra{A}\cq_{\rm sub}\sket{B}$, which must be finite.
If we now consider a sufficiently large collection of matrix elements
(corresponding to various choices of $\sbra{A}$ and/or $\sket{B}$),
we expect that, for each $i$, $\sbra{A}\co^8_i\sket{B}$ will depend
in a different way on all the quark masses. The obvious way for
all the subtracted matrix elements to be finite simultaneously
is, therefore, that the $\b_i^{\rm sub}$ be finite.
As an example of this, the combinations $\b^8_{14}$
and $2(\b^8_3-8\l_8\a^8_2)+2(\b^8_{10}-4\l_5\a^8_2)+\b^8_{11}-(\tilde{\b}^8_1-
8\tilde{\l}_1\a^8_2)$ in Eq.~(\ref{ktopi}) have to be finite.
We thus believe that it is only
a matter of more work to prove that all the $\b_i^{\rm sub}$'s are
in fact finite.

\vspace{1ex}
Our next conjecture deals with the separation of
short- and long-distance effects.

\vspace{1ex}\noindent
{\it Conjecture 2.} The subtraction coefficient $\cf$ defined
in Eq.~(\ref{Kvac}) contains no divergences of the form
$(1/a^2)(B_0 m/f^2)^n$ for $n \ge 1$. In other words,
$\cf_2$ (Eq.~(\ref{Fa})) corresponds only to the $\a^8_2$ term in Eq.~(\ref{Fchpt})
and is independent of all mass parameters, while the finite $\cf_0$
corresponds to the entire NLO part of Eq.~(\ref{Fchpt}).

The evidence supporting this conjecture is that the NLO part
of $\cf$, Eq.~(\ref{Fchpt}), can be expressed as a function of
$\b^{\rm sub}_i$ only. Conjecture~2 thus follows from
Conjecture~1, to NLO.
We believe that this demonstrates in the case at hand
how the short-distance divergences
are disentangled from the long-distance physics.

\vspace{1ex}
In matrix elements, the short-distance divergences get multiplied
by factors that originate from the long-distance physics of the matrix element.
It should therefore come as no surprise that {\it un-subtracted} matrix element
contain divergences of the form $(1/a^2)(B_0 m/f^2)^n$ for $n \ge 1$.

Our results for $\cf$ and $\sbra{\pi^+}\co_{\rm sub}\sket{K^+}$ are
different from those found in ref.~\cite{ls}.%
\footnote{See Eq.~(62) of that paper.}  The reason is that in ref.~\cite{ls}
the subtraction condition Eq.~(\ref{Kvac}) was only imposed to LO in ChPT,
taking $\cf=8\a^8_2/f^2$ instead of Eq.~(\ref{Fchpt}).  This amounts to
using a different subtraction condition, because the NLO
combinations $\b^8_1-8\l_8\a^8_2$, $\tilde{\b}^8_1-8\tilde{\l}^8_1
\a^8_2$ and $\b^8_2-16\l_6\a^8_2$ are
finite (up to logarithms), as we argued above.  Thus, to NLO, the
subtraction condition used in ref.~\cite{ls} is of the form
$\sbra{0}\cq'_{\rm sub}\sket{K^0}=C\ne 0$, with $C$ finite.  Comparing
to Eq.~(\ref{FF'}), indeed $\cf_2=\cf'_2$, but $\cf_0-\cf'_0$ does not vanish,
and is given by the NLO terms of Eq.~(\ref{Fchpt}).

While both subtraction conditions thus remove power divergences
from all matrix elements,
the subtraction condition of Eq.~(\ref{Kvac}) is easier to implement numerically,
because no reference to ChPT is made.  The condition discussed in ref.~\cite{ls}
distinguishes between LO and NLO, and therefore its implementation
requires fitting to the appropriate ChPT expressions.

Finally, we emphasize again that physical results do not depend on which
subtraction condition is chosen,
because $\sbra{\pi^+\pi^-}\co_2\sket{K^0}=0$.  As already mentioned,
at the level of quark operators, this equality follows from the fact that
the parity-odd part of $\co_2$ is a total divergence on-shell;
this argument can be carried over to ChPT using the chiral WT identities.

%%%%%%
\ssec{3.2}{The case $\mres\ne 0$}
%%%%%%
We now investigate how the situation changes when $\mres\ne 0$.
The chiral symmetry breaking, present in lattice QCD with domain-wall
fermions due to the fact that the limit $L_5\to\infty$ is not taken,
can be represented by a local operator transforming in the
same way as an explicit mass term \cite{dwf}.
We take the effect of this chiral symmetry breaking
into account by introducing a new spurion field $\Mres$, which
is then set equal to
\begin{equation}
\Mres={\rm diag}(\mres^v,\dots,\mres^s,\dots,\mres^v,\dots)\,.
\label{Mres}
\end{equation}
$\Mres$ is expected to be small if both valence and sea quarks
are well outside the Aoki phase \cite{lcl}.
The first and last set of entries refer to the valence and
ghost sectors, while the middle set of entries refers to the
sea sector.  Both $\mres^v$ and $\mres^s$ are flavor independent.
(For a more detailed discussion of the systematics of
residual chiral symmetry violations, see Appendix~A.)
We may thus define new building blocks for
operators in the effective theory \cite{newreference}:
\begin{equation}
\Xres_\pm=2B_0(\Sigma\Mres^\dagger\pm\Mres\Sigma^\dagger)\,.
\label{Xres}
\end{equation}
The choice of $2B_0$ as a proportionality factor
defines the normalization of $\mres^{v,s}$.

The presence of $\Mres$ leads to new
$CPS$-even operators in the effective theory:
\begin{eqnarray}
\a^8_2\,\str(\L X_+)&\to&\theta_2\,\str(\L\Xres_+)\,,\label{mresops}\\
\b^8_1\co^8_1&\to&\h_{1,1}\,\str(\L\{X_+,\Xres_+\})+\h_{1,2}\,
\str(\L\Xres_+\Xres_+)\,,\NON
\b^8_2\co^8_2&\to&\h_{2,1}\,\str(\L X_+)\,\str(\Xres_+)
+\h_{2,2}\,\str(\L\Xres_+)\,\str(X_+)\NON
&&\hspace{4.09cm}+\ \h_{2,3}\,\str(\L\Xres_+)
\,\str(\Xres_+)\,,\NON
\b^8_3\co^8_3&\to&\h_{3,1}\,\str(\L\{X_-,\Xres_-\})+
\h_{3,2}\,\str(\L\Xres_-\Xres_-)\,,\NON
\b^8_4\co^8_4&\to&\h_{4,1}\,\str(\L X_-)\,\str(\Xres_-)
+\h_{4,2}\,\str(\L\Xres_-)\,\str(X_-)\NON
&&\hspace{4.09cm}+\ \h_{4,3}\,\str(\L\Xres_-)
\,\str(\Xres_-)\,,\NON
\b^8_{10}\co^8_{10}&\to&\h_{10}\,\str(\L\{\Xres_+,L_\mu L_\mu\})\,,\NON
\b^8_{11}\co^8_{11}&\to&\h_{11}\,\str(\L L_\mu\Xres_+L_\mu)\,,\NON
\b^8_{13}\co^8_{13}&\to&\h_{13}\,\str(\L\Xres_+)\,\str(L_\mu L_\mu)\,,\NON
\b^8_{14}\co^8_{14}&\to&\h_{14}\,\str(\L L_\mu L_\mu)\,\str(\Xres_+)\,,\NON
\b^8_{15}\co^8_{15}&\to&\h_{15}\,\str(\L[\Xres_-,L_\mu L_\mu])\,.\nonumber
\end{eqnarray}
In the strong effective lagrangian, there are new LO and NLO
operators of the form (giving only the ones relevant for the
matrix elements of interest)
\begin{eqnarray}
\str(X_+)&\to&\str(\Xres_+)\,,\label{lmres}\\
\l_4\,\str(L_\mu L_\mu)\,\str(X_+)&\to&\k_4\,\str(L_\mu L_\mu)
\,\str(\Xres_+)\,,\NON
\l_5\,\str(L_\mu L_\mu X_+)&\to&\k_5\,\str(L_\mu L_\mu
\Xres_+)\,,\NON
\l_6\,(\str X_+)^2&\to&\k_{6,1}\,\str(X_+)\,\str(\Xres_+)
+\k_{6,2}\,(\str\Xres_+)^2\,,\NON
\l_7\,(\str X_-)^2&\to&\k_{7,1}\,\str(X_-)\,\str(\Xres_-)
+\k_{7,2}\,(\str\Xres_-)^2\,,\NON
\l_8\,\str(X_+^2+X_-^2)&\to&\k_{8,1}\,\str(X_+\Xres_+
+X_-\Xres_-)+\k_{8,2}\,\str((\Xres_+)^2+(\Xres_-)^2)\,.\nonumber
\end{eqnarray}
The contributions of these new operators
to $\co_2$ can easily be obtained by applying the
chiral rotation $M\to(\L+\L^\dagger)M$, $\Mres\to(\L+\L^\dagger)\Mres$
to these new operators.

First consider the case that only $\mres^s\ne 0$, while  $\mres^v=0$,
\ie\ the valence quarks have the same chiral symmetry as in
the continuum theory.  Clearly, for fixed,
non-zero $\mres^s$, the symmetry group of Eq.~(\ref{pqsym}) gets
broken to the group of Eq.~(\ref{sym}).  Keeping only tree-level
contributions from both LO operators and from the NLO operators
listed in Eq.~(\ref{mresops}), we find that
\begin{eqnarray}
\bra{0}Q\sket{K^0}&=&{4iB_0(m_s-m_d)\over f}
\left(\a^8_2+{4B_0\over(4\pi f)^2}\,\h_{2,1}N_s\mres^s\right)\,,\label{kme}\\
\sbra{\pi^+}Q\sket{K^+}&=&{8B_0m^v\over f^2}\left(\a^8_1
-{4B_0\over(4\pi f)^2}\,\h_{14}N_s\mres^s-\left(\a^8_2+
{4B_0\over(4\pi f)^2}\,\h_{2,1}N_s\mres^s\right)
\right)\,,\NON
&&\label{kme2}
\end{eqnarray}
in which $m^v$ is again the degenerate valence quark mass in the $K^+\to\pi^+$
decay.\footnote{
  We see that all that happens is that,
  in the un-subtracted matrix elements, the LECs $\a^8_1$
  and $\a^8_2$ ``get shifted'' by an amount proportional to $\mres^s$.
  This is a natural consequence of the form of the relevant NLO operators.
  The reader can verify that any ``shift'' in $\a^8_2$
  drops out of the subtracted matrix element.
}
One can show that $\h_{2,1}$ also diverges like $1/a^2$, using
arguments similar to those discussed in Sect.~3.1.
The $\h_{2,1}$ dependent terms are thus new examples of Corollary~2.

The subtraction of power divergences based on Eq.~(\ref{Kvac}) still
goes through unchanged.
Explicitly, we find that
\begin{eqnarray}
\cf&=&\cf\,\Big|_{\mres=0}+{32B_0\over (4\pi f)^2f^2}
(\h_{2,1}-8\k_{6,1}\a^8_2)N_s\mres^s\,,\qquad\label{fmres}\\
\sbra{\p^+}\cq_{\rm sub}\sket{K^+}&=&
\sbra{\p^+}\cq_{\rm sub}\sket{K^+}\bigg|_{\mres=0}
- \;{32B_0^2m^v\over (4\pi f)^2f^2}\, \h_{14}\, N_s\mres^s\,.
\label{kmesub}
\end{eqnarray}
This result shows that $\h_{14}$ is finite.
We also expect that
$\h_{2,1}^{\rm sub}=\h_{2,1}-8\k_{6,1}\a^8_2$ is a new finite
combination of LECs, conform Conjectures 1 and 2.

Of course there are corrections to these results:
Eq.~(\ref{massren}), which expresses the physical pseudo-scalar mass in terms
of chiral-lagrangian parameters at NLO, gets modified
by the $\k$ LECs defined in Eq.~(\ref{lmres}). Also the
chiral logarithms get modified: they are those
of ref.~\cite{gp}, with the proviso that now $M_{SS}^2=2B_0(m^s
+\mres^s)$.  The corresponding $\co_2$ matrix elements
needed for the subtraction get new contributions from the
$\mres^s$ dependence of the strong chiral lagrangian to NLO.
However, none of these corrections
changes the conclusion that the subtraction procedure still works.
The underlying reason is that the operator $\co_2$ in Eq.~(\ref{O4})
is still uniquely the only lower-dimension operator that $\cq$
can mix with.

The situation changes if we now also let $\mres^v$ be non-zero.
In that case, also the valence chiral symmetry in Eq.~(\ref{sym})
gets broken, and another bilinear operator $\co'_2=2\mres^v
\bar{s}d$ appears which can mix with $\cq$ and $\co_2$.
Let us see how this works in ChPT, again omitting
mass and wave-function renormalizations, as well as chiral
logarithms.  Instead of Eqs.~(\ref{kme}) and~(\ref{kme2}) we find
\begin{eqnarray}
\bra{0}\cq\sket{K^0}&=&\bra{0}\cq\sket{K^0}\bigg|_{\mres=0}
\label{kmev}\\
&&+\; {16iB_0^2(m_s-m_d)\over (4\pi f)^2f}
\left(\h_{2,1}N_s\mres^s+2\h_{1,1}\mres^v\right)\,,\NON
\sbra{\pi^+}\cq\sket{K^+}&=&\sbra{\pi^+}\cq\sket{K^+}\bigg|_{\mres=0}
+{8B_0\mres^v\over f^2}\,(\a^8_1-\theta_2)
\label{kmev2}\\
&&-\; {32B_0^2\mres^v\over (4\pi f)^2f^2}
\Biggl((2\b^8_{10}+\b^8_{11})m^v+\b^8_{14}\sum_i m_i^s \Biggr)\NON
&&-\; {32B_0^2(m^v+\mres^v)\over (4\pi f)^2f^2}\Biggl(\h_{14}N_s\mres^s
+(2\h_{10}+\h_{11})\mres^v\Biggr)\NON
&&-\; {32B_0^2m^v\over (4\pi f)^2f^2}\Biggl(\h_{2,1}N_s\mres^s
+4(\h_{1,1}+\h_{3,1})\mres^v\Biggr)\NON
&&-\; {32B_0^2\mres^v\over (4\pi f)^2f^2}
\Biggl(\h_{2,3}N_s\mres^s
+2(\h_{1,2}+\h_{3,2})\mres^v+\h_{2,2}\sum_i m^s_i\Biggr)\,.\nonumber
\end{eqnarray}
The terms involving $\a^8_1$ or $\b^8_i$ arise because the physical
(tree-level)
valence-meson mass gets modified to $M_{phys}^2=2B_0(m^v+\mres^v)$.

The problem one faces when $\mres^v\ne 0$ comes from the fact that
two power-divergent subtractions (both of order $1/a^2$) are
needed now, because of the presence of the two lower-dimension
operators $\co_2$ and $\co'_2$.  An analysis following the lines
of Sect.~3.1 shows that the LECs
$\theta_2$, $\h_{1,i}$, $\h_{2,i}$, $\h_{3,i}$ and $\h_{10}$
all are expected to diverge as $1/a^2$.
Performing only one subtraction with $\co_2$ would still
leave $1/a^2$ divergences in the once-subtracted matrix elements.
The only way out is to define a twice-subtracted operator
\begin{equation}
  \cq_{\rm two-sub}=\cq-\cf\co_2-\cf'\co'_2\,,
\label{twice}
\end{equation}
for which {\it all} matrix elements are finite,
using two independent subtraction conditions.

The terms that depend only on $\mres^v$ or $\mres^s$ in the
$K^+\to\pi^+$ amplitude are in principle not a problem, since they can
be separated out by varying $m^v$ and $m_i^s$, and thus be discarded.
But, due to the divergent LECs $\b^8_{10}$,
$\h_{1,1}$, $\h_{3,1}$ and $\eta_{10}$,
the slope of $\sbra{\pi^+}\cq\sket{K^+}$ as a function of $m^v$,
even after subtracting $\sbra{\pi^+}\co_2\sket{K^+}$ with $\cf$
determined from Eq.~(\ref{Kvac}), remains power divergent.
This magnifies the effect of $\mres^v$, which therefore
will need to be small enough to keep this systematic
error under control.  Of course, the $\mres^s$ dependence
also introduces an error in the determination of $\a^8_1$,
but this error does not get magnified by $1/a^2$ if $\mres^v=0$
(\seef Eq.~(\ref{kmesub})).
The conclusion is that the demand for good chiral symmetry in
the valence sector is much more stringent than in the
sea sector.

As mentioned, the results of Eqs.~(\ref{kmev}) and (\ref{kmev2}) are not complete
without chiral logarithms and mass and wave-function
renormalizations.  But these ``decorations" do not alter
our conclusions on the demand on the smallness
of $\mres^v$.

%%%%%%%%%%%%
\vspace{5ex}
\noindent {\large\bf 4.~Bounds}
\secteq{4}
\vspace{3ex}

In this section we will address the question:
In order that the systematic error due to neglecting $\mres$ effects
will be at the 1\% level, how small should $\mres^v$ and $\mres^s$ be?
Obtaining such bounds requires us to assume something on the magnitude of
$\mres$-dependent contributions, relative to the ``typical'' size
of NLO effects in ChPT.
To account for the possible occurrence of numerically large LECs
we will include a ``safety margin'' of a factor 10.
Namely, we will require that
the magnitude of the neglected terms is 1\promil (0.1\%) when LECs assume
their ``typical'' anticipated magnitude. We will apply similar considerations
to other systematic effects such as (logarithmic) operator mixings
and scaling violations.

%%%%%%
\ssec{4.1}{How small should $\mres^v$ be? (Observation 3)}
%%%%%%
The method of constructing $K\to \pi\pi$ matrix elements from
$K\to \pi$ and $K\to 0$ is highly vulnerable to any non-zero $\mres^v$
because, as we discussed in Sect.~3.2,
 for  $\mres^v\ne 0$, it takes {\it two} subtractions to remove
all the power divergences. Once-subtracted matrix elements will
therefore still contain power-divergent terms.

The magnitude of the un-subtracted $K^+ \to \p^+$ matrix element is
parametrically of order $(1/a^2)(B_0 m/f^2)$.
Here and below, $m$ should be understood as a shorthand for the
linear combination of valence-quark masses relevant for the specific
matrix element. In any given numerical computation, the magnitude of the
subtracted matrix element
can be written as $(\D/a^2)(B_0 m/f^2)$.
Quenched simulations \cite{RBC,CPPACS} with $a^{-1} \sim 2$ GeV find
that $\D$ is a few percent,\footnote{
  This is consistent with $\L_{QCD}^2 \approx 0.01/a^2$.
}
and we will use $\D \approx 1\%$ in our bounds.

Before we derive our bound, let us comment on the situation if
one considers only LO in ChPT.
The LEC $\a^8_1$ is extracted from the slope of the once-subtracted
$K^+ \to \p^+$ matrix element with respect to $m^v$,
while the divergence associated with $\co'_2$
(the $\th_2$ term in Eq.~(\ref{kmev2}))
is given by the intercept. Since the slope is independent
of the intercept, this procedure effectively amounts
to doing both subtractions at the same time
at LO. See \eg Fig.~25 of ref.~\cite{RBC}.\footnote{
  From this figure one deduces that, like $\a^8_2$, the LO LEC $\th_2$
  (Eq.~(\ref{mresops})) is parametrically of order $1/a^2$, as expected.
}
As we showed in Sect.~3.2, this is an artifact of LO ChPT,
and not a valid procedure at NLO.
The fact that terms at NLO compete with LO terms
does not indicate a failure of ChPT. What we have here is a situation
where a {\it divergent} NLO contribution competes with a {\it finite} LO one.
The problem would go away had we made two subtractions!
In that case, the bound on $\mres^v$ would be closer
the bound on $\mres^s$ (discussed in Sect~4.2 below).
We re-examine this issue in Sect.~5.

At $\mres^v\ne 0$, once-subtracted matrix elements will have NLO
power divergences which are parametrically of order
$(1/a^2)(B_0 m/f^2)B_0 \mres^v/(4\p f)^2$. Our bound follows from
the requirement that this will not exceed 1\promil of the physical answer.
Dropping common factors, we obtain
\begin{equation}
  B_0 \mres^v/(4\p f)^2 \le 0.001 \D \approx 10^{-5} \,.
\label{boundv}
\end{equation}
It may be more useful to re-express this bound in terms of
the quantity $a\mres^v$, which is the dimensionless residual mass
measured directly in the simulation. Both $B_0$ and $4\p f$ are roughly
of order 1 GeV, and the same is true for the inverse lattice
spacing used in numerical simulations. Hence, in practice,
$B_0/(4\p f)^2 \approx a$ and we may re-write Eq.~(\ref{boundv})
as $a\mres^v \le 10^{-5}$.

Taking $B_0 m/(4\p f)^2$ as the chiral expansion parameter
is based on the known values of NLO strong LECs.
The LECs $\l_i$ (\seef Eq.~(\ref{lL})) suitable for this parameterization
are indeed $O(1)$.
What would happen if it turns out that, say, $B_0 m/(4\p f^2)$
(rather than $B_0 m/(4\p f)^2$) is the ``true'' chiral expansion parameter
for the relevant weak matrix elements? Since $4\p \approx 10$,
would this change our estimate by the same amount?
The answer is no! It is precisely because of this uncertainty
that we have already included the safety margin of 10 in our estimate.
Thus, requiring that $a\mres^v$ be of order $10^{-6}$ would amount
to a safety margin of a factor 100.

%%%%%%
\ssec{4.2}{How small should $\mres^s$ be? (Observation 4)}
%%%%%%
 From now on we will assume that $a\mres^v \le 10^{-5}$,
and, hence, that we may use all the results derived for $\mres^v=0$.
As for $\mres^s$, we may or may not neglect it the analysis of a
simulation,
and the resulting error will be slightly different in each case.

Let us first assume that $\mres^s$ is neglected, namely,
one performs the ChPT matching at $\mres^v=\mres^s=0$.
The obvious advantage of this choice is that the analysis
involves the minimal number of NLO parameters as dictated by continuum ChPT.
The neglected $\mres^s$-dependent terms are again NLO,
but this time they contain no power divergence, because
the single subtraction of Sect.~2
removes the entire power divergence for $\mres^v=0$.
We already know from the LO analysis that the magnitude
of a ``typical'' finite LEC is $\D$ times the magnitude of a ``typical''
power-divergent LEC. The neglected terms are therefore of order
$(\D/a^2)(B_0 m/f^2)B_0 \mres^s/(4\p f)^2$,
and this should be less than 1\promil times the LO finite result.
Dropping all common factors (including $\D$ this time)
gives rise to the simple result
\begin{equation}
  B_0 \mres^s/(4\p f)^2  \le 0.001 \,,
\label{bounds}
\end{equation}
As before, this bound is the same as
$a\mres^s \le 0.001$ if $a^{-1}\sim 1$~GeV.

Compared to Eq.~(\ref{boundv}), the bound~(\ref{bounds}) is far less stringent.
We now consider also short-distance
systematic errors induced by neglecting $\mres^s$.
This includes mixings between dimension-six operators that are allowed
by the symmetry of Eq.~(\ref{sym}) but not by the full PQ symmetry~(\ref{pqsym}).
Operator mixings are determined (\eg using the RI NPR method \cite{RBC})
at off-shell momenta of roughly $p \approx 1/a$. The resulting
``wrong-chirality'' operator mixing will therefore be at most of order
$a\mres^s$, and requiring this to be a 1\promil effect
(safety margin included) we again end up with $a\mres^s \le 0.001$.\footnote{
  This bound is obtained directly in terms of $a\mres^s$
  without using the ``translation'' $B_0/(4\p f)^2 \approx a$.
}

We next consider what scaling violations are induced at non-zero $\mres^s$.
At $\mres^v=\mres^s=0$, domain-wall fermions have only $O(a^2)$
scaling violations. Since $a\Lambda_{QCD} \approx 0.1$,
this is a 1\% effect. For $\mres^s \ne 0$ we may have $O(a)$
scaling violations coming from the dimension-five operator
$\overline{q} \s_{\m\n} F_{\m\n} q$. Exact chiral symmetry forbids
this operator, hence it can at most arise with a coefficient of order
$a\mres^s$. Requiring that this effect will be smaller than
$0.1(a\Lambda_{QCD})^2$ we end up with the same bound as before.

In summary, taking known sources of systematic error into account
we conclude that, provided $a\mres^s \le 0.001$,
all $\mres^s$ effects can be completely neglected and the resulting
error due to $\mres^s$ should be at the 1\% level.

Of course, several independent 1\% errors might accumulate
to a somewhat bigger error. If one does ChPT at non-zero $\mres^s$,
the systematic error generated by the ChPT matching
will be entirely due to NNLO effects.
However, this would require more measurements to extract
the new linear combinations of NLO LECs that occur in
$K^0 \to \p^+\p^-$ when $\mres^s\ne 0$.

We anticipate that the numerical calculation of dimension-six operator mixing
should be feasible assuming only the symmetry of Eq.~(\ref{sym})
which takes into account that $\mres^s \ne 0$. (See \eg the NPR mixing matrix
in Tables 12,13 of ref.~\cite{RBC}, where mixings that violate
$SU(3)_L\times SU(3)_R$ are not neglected.)

Quark masses commonly used in numerical simulations are typically in the range
of few times 10 MeV, and so $a\mres^s \approx 0.001$ may already
account for up to 10\% of the physical sea-quark mass.
We thus believe that, no matter what, one should avoid residual masses
(be it $\mres^s$ or, certainly, $\mres^v$) larger than 0.001.

The good chiral symmetry of domain-wall fermions
enhances a quenched pathology: due to the absence of the fermion determinant,
contributions of physical zero modes grow like the inverse quark mass.
While methods have been devised to deal with this problem in
the quenched theory \cite{RBC2}, clearly the only true solution
is to un-quench. The pathology is removed when
the physical sea and valence masses are set to the same value.
To LO, this means
\begin{equation}
  m^{\rm ph}_i = m_i^v+\mres^v = m_i^s+\mres^s \,.
\label{mph}
\end{equation}
Observe that this does {\it not} require that $\mres^v$ and $\mres^s$ be
equal. When $\mres^v$ is negligibly small, this amounts to tuning
the valence-quark mass $m_i^v$ to $m_i^s+\mres^s$. In a numerical
simulation, the correct procedure is to tune the valence-pion mass
to the sea-pion mass. Another advantage
of this choice is that enhancement of finite-volume effects is avoided
\cite{ls}.

%%%%%%%%%%%%
\vspace{5ex}
\noindent {\large\bf 5.~Discussion}
\secteq{5}
\vspace{3ex}

Our analysis shows that, because power divergences are encountered in
the valence sector only, the chiral-symmetry requirements
on valence quarks are much more stringent in comparison to sea quarks.
To keep systematic errors due to chiral symmetry violation
at the 1\% level, we estimate that one needs $a\mres^v \le 10^{-5}$
and  $a\mres^s \le 10^{-3}$ at $a^{-1}\sim 2$~GeV. The obvious conclusion
is to adopt the strategy of using different DWF actions
for the two sectors. A more economic DWF action can, and should,
be chosen for the sea sector; it is the sea sector that costs most
of the computational effort, and at the same time this is
where a much larger residual mass can be tolerated!

The simplest way to choose different DWF actions for the two sectors
is to take $L^s_5< L^v_5$.
This choice is a hybrid scheme in the sense of
ref.~\cite{brs}, with symmetry group $G_{sea}\times G_{valence}=U(N_s)
\times U(N_v|N_v)$.  For $L^s_5=L^v_5$ the symmetry group enlarges to
the
group $U(N_v+N_s|N_v)$ \cite{bg}, and in that case
the theory is what is usually referred to as partially quenched,
rather than hybrid.  While these are all rather trivial statements,
they have one relevant consequence: The most general theory, with
free parameters $m_q^{v,s}$ and $L^{v,s}_5$, rigorously contains fully
unquenched
QCD as a special case, by choosing $m_q^v=m_q^s$, $L^v_5=L^s_5$.
However, while simulations can easily obtain data points for
$m_q^v=m_q^s$, this is less likely the case for the size of the fifth
dimension. The problem is that, as we have seen,
choosing $L^v_5=L^s_5$ for affordable values of $L^s_5$ is unlikely
to yield the high chiral accuracy needed in the valence sector.
Note that one might choose hybrid schemes in which the valence-
and the sea-quark DWF operators differ not only in the values
of $L_5$ or $m_q$, but also in other ways (see below).

Of course, spectral functions extracted from
the mixed theory, in which quark masses have been tuned such that
Eq.~(\ref{mph}) holds, may suffer from violations of positivity at non-zero
lattice spacing.
However, provided the bounds discussed in Sect.~4 are satisfied,
implying that $a(\mres^s-\mres^v)$ errors are not bigger than $a^2$ errors,
this will show up as just another source of $O(a^2)$ scaling violations.
If Eq.~(\ref{mph}) does not hold, one expects unitarity violations to
occur also in the continuum limit as in any partially-quenched theory.

In the quenched case, changing the gauge action
(to Iwasaki \cite{Iw} or to DBW2 \cite{DBW2})
turned out to be enough to push $\mres$ to negligibly small values.
In the dynamical DWF case \cite{ddwf}, fiddling only with the gauge action
gives rise to $\mres=\mres^s$ values which
at present are marginally within our bound
for two flavors, and larger than our bound for three flavors. Moreover,
trying to use exactly the same DWF action for the valence sector would
entail violating our $\mres^v$ bound by several orders of magnitude.

The inherent reason for the larger $\mres$ values obtained in
the dynamical case is the screening effect of the sea quarks,
requiring a larger bare coupling in order to maintain
the same lattice cutoff. This larger bare coupling pushes one
closer to the Aoki phase \cite{aoki,shsi,scri}. Inside the Aoki phase,
DWF and overlap fermions \cite{ovlp} cannot be used
because it is impossible to maintain both chirality and locality
\cite{hjl,lcl}.

To meet the bounds derived in this paper it is necessary
to use better DWF actions (for both fermions and Pauli-Villars fields).
We recall that the basic elements that go into a DWF construction
are the four-dimensional ``kernel''
(for standard DWF, a super-critical Wilson operator)
and the way the fermions are allowed to hop in the fifth dimension
(for standard DWF, a free one-dimensional Wilson operator).
Both play a role in the resulting Ginsparg-Wilson (GW)
operator \cite{gw} obtained in the chiral limit $L_5\to\infty$,
and faster approach to the DWF chiral limit often gives rise
to a GW operator with better locality properties.
While the study of improved DWF actions is outside the scope
of this paper, we briefly list several existing proposals.

\vspace{1ex}\noindent
{\it 1.} Accelerated convergence to the chiral limit \cite{abc}.
This method changes the couplings in the fifth dimension,
in effect applying a M\"obius transformation to the Wilson kernel.
It includes as special cases the standard DWF formulation \cite{dwf}
and the ``truncated overlap'' of ref.~\cite{AB}. It shows promising results.
It can be applied to both the sea- and the valence-quark sectors.
The difference between the sea and the valence DWF actions can
come not only from ($m$ and) $L_5$ but also from two optimization parameters,
see ref.~\cite{abc} for details.

\vspace{1ex}\noindent
{\it 2.} Modified four-dimensional kernel. Modifications have
been proposed based on the rate of approach to the chiral limit
in perturbation theory \cite{next},
and on approximate solutions \cite{wb} of the GW relation
which also improve the locality of the resulting overlap operator.
These methods are again applicable to both sea and valence quarks.

\vspace{1ex}\noindent
{\it 3.} Modified Pauli-Villars action \cite{better}.
This method is relevant only to the sea-quark sector.
It aims to suppress the Boltzmann weight of dislocations that give rise to
low-lying eigenmodes of the Wilson kernel.
These eigenmodes allows for un-suppressed propagation in the fifth dimension,
and are the dominant non-perturbative source of chiral symmetry violation.

\vspace{1ex}\noindent
{\it 4.} ``Projection method'' \cite{jn}.
In this method, one adds to the DWF Dirac operator terms proportional
to the projectors on low-lying eigenmodes of the Wilson kernel.
This in effect raises the corresponding eigenvalues and eliminates
the associated un-suppressed propagation in the fifth dimension.
The method is consistent so long as these eigenmodes are exponentially
localized. The method could be prohibitively expensive for the sea sector.
As for the valence sector, the extra cost of computing the projection terms
should be small compared to the cost of generating the dynamical
configuration itself. Under favorable conditions, ``projecting out''
a relatively small number of Wilson eigenmodes can lead to a significant
reduction of $\mres^v$.

\vspace{1ex}\noindent
{\it 5.} Another method which would be prohibitively expensive for
the sea sector, but can be used in the valence sector,
is to smear the gauge links before plugging them into the DWF
Dirac operator \cite{stagDWF}. Again, this removes many of the dislocations
that would otherwise give rise to low-lying eigenmodes of the Wilson kernel.

\vspace{1ex}
Once dynamical DWF configurations with $a\mres^s\le 10^{-3}$ have
been produced there is a good chance that, with all the extra
``tricks'' available in the valence sector, it will be possible
to meet the valence-sector bound $a\mres^v\le 10^{-5}$ as well,
at affordable cost.

In Sect.~3 we concluded that, when $\mres^v=0$,
the divergent part of $\cf$ is  $\cf_2/a^2=8\a_2^8/f^2$.
Similarly, if we consider the twice-subtracted operator (Eq.~(\ref{twice}))
needed when $\mres^v\ne 0$, we have that $\cf_2/a^2=8\a_2^8/f^2$
as before, while $\cf'_2/a^2=8\th_2/f^2$. This follows from the LO
analysis, but should remain true to NLO if indeed the NLO corrections
to $\cf$ and $\cf'$ are always finite (\seef Conjecture 2). This suggests that
a twice-subtracted operator may be defined by first performing
one subtraction by imposing Eq.~(\ref{Kvac}), and then performing
the second subtraction, $\cq_{\rm two-sub}=\cq_{\rm sub} -\cf'\co'_2$,
by imposing $\sbra{\p^+}\cq_{\rm two-sub}\sket{K^+}=0$ in the limit
that the explicit valence quark mass is zero. The ChPT matching should
then involve the corresponding NLO result $\cf'=8\th_2/f^2+O(\mres,m^s)$,
where the NLO terms are again finite.
Observe that, unlike the first subtraction, the second subtraction
cannot be done without relying on ChPT to some order.
This brings in new uncertainties (for instance, the convergence of the
chiral extrapolation).
Given these uncertainties,
we believe that the necessary bound
on $a\mres^v$ will fall somewhere in between $10^{-5}$ and $10^{-3}$
in that case.

Finally, it is interesting to see whether the considerations on the interplay
of power divergences and chiral symmetry presented in this paper
carry over to other fermion methods.  For instance, as an alternative
to using DWF, efforts are under way to
use staggered valence quarks to determine $K\to\pi\pi$
from $K\to\pi$ matrix elements \cite{wlee}.  Naively,
one expects that bounds on the
maximum tolerable Goldstone-boson mass splittings coming from
taste symmetry breaking would have to be similarly stringent
as those on $\mres^v$.  The reason is that taste symmetry
breaking in the valence sector might again lead to the need to
perform more than one power-like subtraction.  However, in this
case the problem is avoided by using only the exact
Goldstone bosons on the external lines of the matrix elements
under consideration.

%%%%%%%%%%%%
\vspace{5ex}
\noindent {\bf Acknowledgements}
\vspace{3ex}

We started this work at the workshop ``Matching light quarks to
hadrons" at the Benasque Center for Science in Spain.   We would like to
thank the participants of this workshop,
in particular Oliver B\"ar, Claude Bernard, Jack Laiho,
Weonjong Lee, Bob Mawhinney, Kostas Orginos and Amarjit Soni,
for extensive discussions.
We have also benefited from discussions longer ago with
participants of the program ``Lattice QCD and hadron phenomenology"
at the Institute for Nuclear Theory at the University of
Washington, in particular Steve Sharpe.
MG is supported in part by the US Department of Energy, and
YS is supported by the Israel Science Foundation under grant
222/02-1.

%%%%%%%%%%%%
\vspace{5ex}
\noindent {\large\bf Appendix A. Systematics of
chiral symmetry violations}
\secteq{A}
\vspace{3ex}

In this paper, we assumed that finite-$L_5$ chiral symmetry violations
can be treated in ChPT by introducing a spurion field associated with
the ``residual mass.'' While the bare quark mass
is a small parameter explicitly present in the DWF action,
the same is not true for the residual mass; its smallness is a dynamical
result. In this Appendix we discuss this dynamics in some detail, finding
that a systematic treatment of residual chiral symmetry violations requires
several spurions. However, this will not modify the NLO analysis of
Sect.~3 (except for an inessential modification to Eq.~(\ref{kmev2}))
and, thus, it has no effect on our conclusions.

At finite $L_5$, the DWF's PCAC relation takes the form \cite{dwf}
\begin{equation}
  \partial^-_\m J_{5\m}^a = 2 m J_5^a + 2 J_{5q}^a \,,
\label{pcac}
\end{equation}
where for simplicity we assume equal quark masses.
Here $J_{5\m}^a$ is a DWF's partially-conserved non-singlet axial current
and $\partial^-_\m$ is the backward lattice derivative.
$J_5^a$ is built out of fermion field residing
on the boundaries, and interpolates a (continuum) pseudo-scalar density.
$J_{5q}^a$ is a lattice artifact that involves fermion fields
residing inside the five-dimensional ``bulk,'' half-way between the boundaries.
The notion of a residual mass comes from the fact that the leading
long-distance effect of $J_{5q}^a$ can be expressed as
$J_{5q}^a \approx \mres\, J_5^a$. This motivated the definition
of the spurion $M_{\rm res}$ in Eq.~(\ref{Mres}). In this appendix we address
the legitimacy of this approximation.

Propagation in the fifth dimension is governed by a transfer matrix $T$, from
which a hamiltonian
$\tH = -\log(T)/a_5$ may be defined, where $a_5$ is the
spacing in the fifth dimension. $\tH$ is closely related
to the Wilson operator, and in particular the two operators share
identical zero modes. Each eigenmode of $\tH$
can be characterized as extended or localized. Outside the Aoki phase,
one expects the localized eigenmodes to lie below a mobility edge $\tl_c>0$,
above which the eigenmodes are extended. Following ref.~\cite{lcl}
we will now separately discuss the chiral symmetry violations arising
from the extended and from the localized modes, in that order.

In the presence of extended modes only, low-energy quark modes will
have a factorized wave function
$\j_{DWF}(x,s) \to q_R(x)\c(s) + q_L(x)\c(L_5+1-s)$
where $s=1,2,\ldots,L_5$ is the fifth coordinate
and $q_{R,L}(x) = \half(1\pm\g_5)q(x)$
is an effective four-dimensional quark field. This was derived in
perturbation theory \cite{next}, and is expected to hold
non-perturbatively as well.
The fifth-coordinate wave function $\c(s)$ is dominated by the extended
modes near the mobility edge,\footnote{
  In perturbation theory the mobility edge is equal to the gap
  of the free $\tH$. See also the last paper of ref.~\cite{lcl}.
}
and is given by $\c(s) \sim \exp(-a_5 s\tl_c)$.
Since the left-handed and right-handed quarks
reside on opposite boundaries, each chirality flip
will involve a factor of $\c(L_5) \sim \exp(-a_5 L_5\tl_c)$ when $m=0$.
This resembles a mass insertion with $\c(L_5)$ as the mass.
Hence we can account for it by introducing a mass-spurion $\Mext$
with expectation value
\begin{equation}
\Mext={\rm diag}(\mext^v,\dots,\mext^s,\dots,\mext^v,\dots)\,,
\label{MresE}
\end{equation}
where
\begin{equation}
  a\mext^v \sim \exp(-a_5 L^v_5\tl^v_c)\,,
\label{mresE}
\end{equation}
and similarly for $\mext^s$.

We next turn to the localized modes. Among these, the most dangerous are
the zero modes of $\tH$, which give rise to un-suppressed propagation
between the two boundaries. At a given $L_5$, the term ``zero modes'' stands
here for all modes with eigenvalue $|\l| \leqx 1/L_5$.
The probability (per unit volume) to encounter a zero mode
is roughly given by $\tilde\r(0)/L_5$,
where $\tilde\r(\l)$ is the spectral density
of $\tH$, and $1/L_5$ accounts for the energy interval.\footnote{
  This corresponds to the third scenario described in appendix~C.2
  of ref.~\cite{next}.}
Evidently, the smallness of this effect
hinges on having a small $\tilde\r(0)$. As before,
we introduce a new mass-spurion $\Mlcl$ with
\begin{equation}
\Mlcl={\rm diag}(\mlcl^v,\dots,\mlcl^s,\dots,\mlcl^v,\dots)\,,
\label{MresL}
\end{equation}
in which
\begin{equation}
  a\mlcl^v \sim a^4\tilde\r^v(0)/L^v_5\,,
\label{mresL}
\end{equation}
etc. However, for the chiral symmetry violations coming from localized modes,
this is not the whole story. Once a zero mode is found
some place in the configuration, it allows for quarks of any flavor
to propagate from one boundary to the other with no ``penalty.''
Since the process is confined to the space-time region occupied by
the zero mode, we anticipate that it can be described
in terms of a local effective multi-fermion interaction.
Given \eg $N_s$ sea quarks, a single zero mode can certainly give
rise to effective vertices containing 2,4,$\ldots$, up to $2N_s$
fermions, which induce one chirality flip per flavor.\footnote{
  This might be thought of as reminiscent of the 't Hooft vertices induced
  by instanton zero modes. We leave open how many
  chirality flips can be induced by a single zero mode of $\tH$
  in the case at hand.
}

The Symanzik action is a continuum effective action in which the lattice
artifacts occur in an expansion in the lattice spacing.
In this effective action the multi-fermion vertices discussed above will
have dimensionful coupling constants, each of which will be roughly equal to
the common factor $\tilde\r(0)/L_5$,
times an appropriate power of the lattice spacing.
To account for this effect in ChPT \cite{shsi,brs},
we thus need to introduce a third spurion $W$ transforming again
in the same way as a mass spurion, and having an expectation value
\begin{equation}
W={\rm diag}(w,\dots,w,\dots,w,\dots)\,,
\label{Ahat}
\end{equation}
with
\begin{equation}
  w \sim a^3\,.
\label{ahat}
\end{equation}
Thus, \eg a sea-sector $2n$-fermion vertex (which induces $n$ chirality flips)
will occur at order $\mlcl^s\, a^{3(n-1)}$ in the
chiral expansion.

We end up with the following two rules for residual chiral symmetry
violations in ChPT:
\begin{enumerate}

\item The effects are accounted for by the three mass spurions
$\Mext$, $\Mlcl$ and $W$.

\item The spurion $W$ cannot occur except in terms that already
contain at least one power of $\Mlcl$.
(More precisely, this is true separately for the valence and sea sectors.)
\end{enumerate}

The values of all three spurions are to be defined in the limit
in which the explicit quark mass has been set equal to zero.\footnote{
  One should be able to extract $\mext$ and $\mlcl$ from
  a fit of $\mres$ as a function of $L_5$.
}
ChPT will then automatically take care of the dependence of
the residual mass on the explicit quark masses.
Numerically, for the range of quark masses used in simulations,
the variation in the residual mass
is found to be at the level of a few percent \cite{Iw,DBW2},
showing that it is indeed
consistent to treat it as a higher order effect in ChPT.

For small $L_5$ the dominant
chiral symmetry violating effects will come from extended modes.
Since this effect is damped exponentially, above some value
of $L_5$ the localized zero modes will take over
as the dominant source of chiral symmetry violations.
Their effects have a lot in common with
the chiral symmetry violations of Wilson fermions.
The crucial difference is that, for DWF, these effects have a low probability
given by $\tilde\r(0)/L_5$. This is the origin of Rule~2.

It has been suggested that the treatment of
DWF chiral symmetry violations in ChPT can be based on
promoting the fifth-direction links located exactly in the middle
of the fifth dimension to a spurion $\O$ \cite{newreference}.
However, adopting the more elaborate scheme of spurions developed above
allows one to account more faithfully for the magnitude of residual chiral
symmetry violations coming from the different dynamical sources.

Were it not for $W$, we could define $\Mres=\Mext+\Mlcl$
and the ChPT treatment would be reduced to the single $\Mres$ spurion
introduced in Eq.~(\ref{Mres}). While in general this is not the case,
thanks to Rule~2 it remains true that
the leading residual-mass dependence must be through the sum $\Mext+\Mlcl$.
Let us now re-examine the results of Sect.~3. First, for $\mres^v=0$
(\ie\ for $\mext^v=\mlcl^v=0$), the results depend linearly on
$\mres^s$. The distinction between
the contributions of extended and localized modes therefore plays no
role to NLO when $\mres^v=0$. To follow the above rules,
we would only need to perform the trivial substitution
$\mres^s = \mext^s+\mlcl^s$

When $\mlcl^v$ is non-zero things get (yet) more complicated
compared to Eq.~(\ref{kmev2}). This equation is quadratic in $\mres^v$.
Therefore, apart from setting $\mres^v=\mext^v+\mlcl^v$,
there are additional NLO terms proportional to $\mlcl^v\, a^3$.
This, however, has no effect on the bound~(\ref{boundv}), which originates
from the terms in Eq.~(\ref{kmev2}) proportional to $\mres^v\, m^v$.

So far, we have encountered no difference between the way the
explicit-mass spurion $M$ and the new spurion $\Mext$ enter ChPT.
This raises the question whether
one can trade $M$ and $\Mext$ for a single spurion $M'=M+\Mext$.
The answer depends on the context. For pure-QCD observables (such
as spectrum), the only distinction
is that the LECs corresponding to $M$ and to $\Mext$
will in general exhibit different scaling violations. This is best seen by
considering the effective four-dimensional lattice Dirac operator obtained by
integrating out the five-dimensional bulk modes \cite{eff4}.\footnote{
Note that localized modes can, in principle, be avoided by imposing
an admissibility condition \cite{hjl}.}
This Dirac operator will involve a new mass term with
coefficient $\mext\sim\exp(-a_5 L_5 \tl_c)$. However, in addition
it will contain irrelevant terms transforming in the same way
as a mass term, with coupling constants of order $\exp(-a_5 L_5 \tl_c)$.
Since there are no corresponding irrelevant interactions
for the explicit mass term, the resulting scaling violations
will be different.

However, when insertions of the effective weak hamiltonian are considered,
one encounters the power divergences which are the main topic of this
paper. In this case, the numerical results \cite{RBC,CPPACS}
show that the difference between the LECs corresponding
to $m^v$ and to $\mres^v$, such as for example $\a^8_2$ and $\th_2$,
is roughly of the same size as these LECs.
Therefore it is not possible to combine the explicit-mass
spurion $M$ with any of the residual-mass spurions.

%%%%%%%%%%%%

\end{document}